%% file: MOS_HotSpot_ST_Qubit.tex
\documentclass[prl,aps,twocolumn,showpacs,superscriptaddress]{revtex4-1}

\usepackage{amssymb,amsmath}
\usepackage{graphicx}
\usepackage{multirow}
\usepackage{color}
\usepackage[english]{babel}
\usepackage{hyperref}

\newcommand{\ket}[1]{ |{#1} \rangle }

\begin{document}

\title{\textbf{A silicon singlet-triplet qubit driven by spin-valley coupling}}

\author{Ryan M. Jock}
\email[Corresponding author: ]{rmjock@sandia.gov}
\affiliation{Sandia National Laboratories, Albuquerque, NM 87185, USA}
\author{N. Tobias Jacobson}
\affiliation{Center for Computing Research, Sandia National Laboratories, Albuquerque, NM 87185, USA}
\author{Martin Rudolph}
\affiliation{Sandia National Laboratories, Albuquerque, NM 87185, USA}
\author{Daniel R. Ward}
\email[Present Address: ]{HRL Laboratories, LLC, Malibu, CA 90265}
\affiliation{Sandia National Laboratories, Albuquerque, NM 87185, USA}
\author{Malcolm S. Carroll}
\email[Present Address: ]{Princeton Plasma Physics Laboratory, Princeton, NJ 08543}
\affiliation{Sandia National Laboratories, Albuquerque, NM 87185, USA}
\author{Dwight R. Luhman}
\affiliation{Sandia National Laboratories, Albuquerque, NM 87185, USA}

\begin{abstract}
	
Spin-orbit effects, inherent to electrons confined in quantum dots at a silicon heterointerface, provide a means to control electron spin qubits without the added complexity of on-chip, nanofabricated micromagnets or nearby coplanar striplines. Here, we demonstrate a novel singlet-triplet qubit operating mode that can drive qubit evolution at frequencies in excess of 200 MHz. This approach offers a means to electrically turn on and off fast control, while providing high logic gate orthogonality and long qubit dephasing times. We utilize this operational mode for dynamical decoupling experiments to probe the charge noise power spectrum in a silicon metal-oxide-semiconductor double quantum dot. 
In addition, we assess qubit frequency drift over longer timescales to capture low-frequency noise. We present the charge noise power spectral density up to 3 MHz, which exhibits a $1/f^{\alpha}$ dependence consistent with $\alpha \sim 0.7$, over 9 orders of magnitude in noise frequency.
\\\\

\end{abstract}

\maketitle

\section{Introduction}
\label{sec:Introduction}
\input{sec_Introduction/Introduction.tex} 

\section{Results}
\label{sec:Results}
\input{sec_Device/Device.tex} 
\input{sec_Measurements/Measurements_ElectricalControl.tex} 
\input{sec_Measurements/Measurements_HotSpotQubit.tex} 

\input{sec_Measurements/Measurements_ChargeNoise.tex} 

\section{Discussion}
\label{sec:Discussion}
\input{sec_Discussion/Discussion.tex} 

\section{Methods}
\label{sec:methods}
\input{sec_Methods/Methods.tex} 
\section{Acknowledgments}
\label{sec:Acknowledgments}	

Sandia National Laboratories is a multi-mission laboratory managed and operated by National Technology and Engineering Solutions of Sandia, LLC., a wholly owned subsidiary of Honeywell International, Inc., for the U.S. Department of Energy's National Nuclear Security Administration under contract DE-NA-0003525. This paper describes objective technical results and analysis. Any subjective views or opinions that might be expressed in the paper do not necessarily represent the views of the U.S. Department of Energy or the United States Government.

\section{Author contributions}
\label{sec:contributions}
 R.M.J. performed the central measurements presented in this work and analyzed results. N.T.J. carried out the theoretical modeling and statistical analysis of measurement data. M.R. performed the initial measurements demonstrating the electrical control of valley hot spot rotations on a similar device. All authors discussed central results throughout the project. D.R.W. and M.S.C. designed the process flow, fabricated devices, and designed/characterized the Si material growth for this work. M.S.C. and D.R.L. supervised the combined effort, including coordinating fabrication and identifying modeling needs. R.M.J. and N.T.J. wrote the manuscript with input from co-authors.

\section{Data availability}
\label{sec:DataAvailablity}
The authors declare that the data supporting the findings of this study are available within the paper and its Supplementary Information. Additional data (e.g., source data for figures) are available from the corresponding author upon reasonable request.

\bibliography{bibliography}
\bibliographystyle{unsrt}

\cleardoublepage

\section{Supplementary Information}

\appendix

\input{sec_Appendix/Appendix.tex}

\end{document}

%% file: sec_Introduction/Introduction.tex
Qubits based on the spins of electrons confined to gate-defined quantum dots (QDs) in silicon metal-oxide-semiconductor (MOS) structures have developed into a promising platform for quantum information processing. 
High-quality single-qubit \cite{Veldhorst2014,Yang2019}  and two-qubit gates \cite{Veldhorst2015aa,Huang2019,Yang2019b} have been demonstrated, and device manufacture is generally compatible with available silicon microelectronics fabrication methods.
Qubit control techniques demonstrated in silicon MOS have utilized electron spin resonance (ESR) with microwave strip-lines \cite{Veldhorst2014,Yang2019,Petit2019}, electric dipole spin resonance (EDSR) using micromagnets \cite{Yang2019b} or the intrinsic spin-orbit coupling (SOC) at the Si/SiO$_2$ interface \cite{Jock2018,Corna2018,Harvey-Collard2019,Corna2018}.
Making use of interfacial SOC has the appeal of driving qubit evolution with electrical-only control without reliance on the added fabrication constraints of micromagnets or on-chip microwave strip-lines.

Confining electrons to quantum dots at the Si/SiO$_2$ interface has been shown to produce spin-orbit coupling that is stronger than that of bulk Si \cite{Veldhorst2015,Jock2018,Harvey-Collard2019,Ferdous2018,Tanttu2019}. Recent observations have demonstrated that the broken crystal symmetry at the silicon heterointerface and interactions with excited valley states lead to this enhanced SOC. These effects contribute to variation of the $g$-factor in QDs \cite{Veldhorst2015,Jock2018,Ferdous2018,Fogarty2018,Tanttu2019,Harvey-Collard2019}. 
The $g$-factor difference between neighboring QDs has proven to be a valuable resource, 
able to drive the evolution of spin qubits encoded into a singlet-triplet subspace \cite{Jock2018,Harvey-Collard2019,Jirovec2021}. 
Spin-valley coupling  is known to enhance electron spin relaxation (shorter spin T$_1$) near the ``hot spot'' when the valley splitting, $\Delta_{\mathrm{v}}$, is comparable to the electronic Zeeman splitting, $E_{Z} = g \mu_{\mathrm{B}} B$, in a QD, where $\mu_{\mathrm{B}}$ is the Bohr magneton and $B$ is the applied external magnetic field \cite{Yang2013,Borjans2019,Hollmann2019,Zhang2020}. This enhanced relaxation mechanism has been used to study valley splitting \cite{Yang2013,Hollmann2019,Zhang2020} and intervalley spin-orbit coupling in silicon QD devices \cite{Yang2013,Hao2014,Zhang2020}. Additionally, spin-valley coupling has been proposed as a mechanism to coherently control electron spins in silicon QDs\cite{Huang2017,Bourdet2018,Bourdet2018b}. This could potentially provide an intrinsic qubit control mechanism without the added fabrication complexity of integrated features such microwave striplines and micromagnets. However, coherent qubit control using spin-valley coupling has yet to be experimentally demonstrated in a silicon spin qubit.

In this work, we utilize the intervalley spin orbit interaction near the spin-valley hot spot in a silicon MOS QD and demonstrate the ability to drive singlet-triplet rotations in excess of 200 MHz using the intervalley spin-orbit interaction. We exploit these fast rotations near the hot spot to enable unique qubit operation with high-speed all-electrical modulation between qubit logic gates and high orthogonality of control axes through electrical control of the valley splitting. These fundamental measurements establish this qubit as a candidate for future quantum information processing systems.
We take advantage of this novel operating mode to investigate the charge noise power spectral density (PSD) in this device. We use the noise filtering properties of a Carr–Purcell–Meiboom–Gill (CPMG) dynamical decoupling pulse sequence to decouple the qubit from charge noise experienced during the spin-spin exchange interaction. Combined with long-timescale measurements of drift in the frequency of exchange driven ST qubit rotations, we find that charge noise in this device exhibits a power spectral density consistent with $S(f) \sim f^{-0.7}$ over 9 decades of frequency.

%% file: sec_Device/Device.tex
\subsection{\textbf{Intervalley spin-orbit interaction}}

\begin{figure*}
	\centering
	\includegraphics[width=0.95\textwidth]{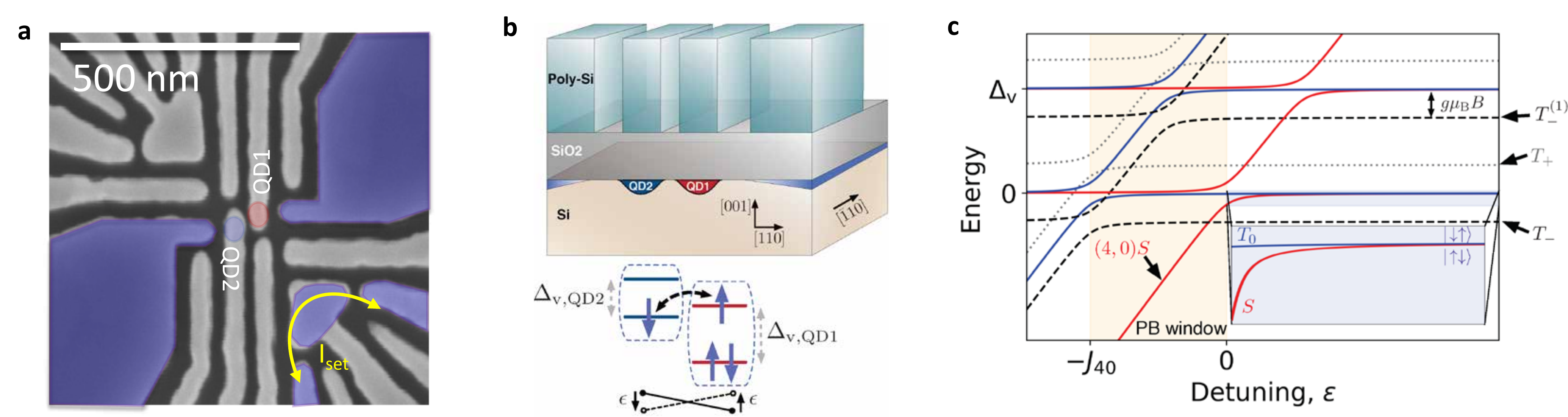}
	\caption{MOS DQD singlet-triplet qubit device.(a) Scanning electron micrograph of the gate structure of a device similar to that measured. The shaded regions indicate estimated areas of electron accumulation. The red and blue circles represent the locations of QD1 and QD2, respectively. We sense  QD charge state transitions using a nearby single electron transistor (SET) in the lower right corner. (b) Schematic lateral view of the device structure and representation of the electron spin filling in each QD. (c) Energy level diagram of the singlet-triplet system in the DQD. The orange region represents the Pauli blockade window, with the singlet-triplet splitting in the (N$_{\textrm{QD1}}$,N$_{\textrm{QD2}}$) = (4,0) charge region denoted by $J_{40}$. (inset) Energy level diagram of the $m$ = 0 qubit subspace in the (N$_{\textrm{QD1}}$,N$_{\textrm{QD2}}$) = (3,1) charge region.}
	\label{fig:Device}
\end{figure*}

The silicon MOS double-quantum dot (DQD) used in this work is illustrated in Fig. \ref{fig:Device}(a). We operate the device near the (N$_{\textrm{QD1}}$,N$_{\textrm{QD2}}$) = (4,0)-(3,1) charge transition. Two electrons on QD1 form a spin paired closed shell \cite{Higginbotham2014,Harvey-Collard2017,Leon2019}. The interaction between the remaining two electrons is electrically controlled via the detuning bias, $\epsilon$, between the QDs. For shallow detuning, there is significant electronic wave function overlap between the two electrons and the exchange energy, $J(\epsilon)$, is the dominant interaction. When the two electrons are well separated in the deep tuning regime, $J(\epsilon)$ is small and the dominant interaction is set by the interfacial SOC, which results in distinct Zeeman energies in each QD~\cite{Jock2018,Jones2018,Fogarty2018,Ferdous2018,Harvey-Collard2019,Tanttu2019}. 

%% file: sec_Measurements/Measurements_ElectricalControl.tex
\begin{figure*}
	\centering
	\includegraphics[width=0.99\textwidth]{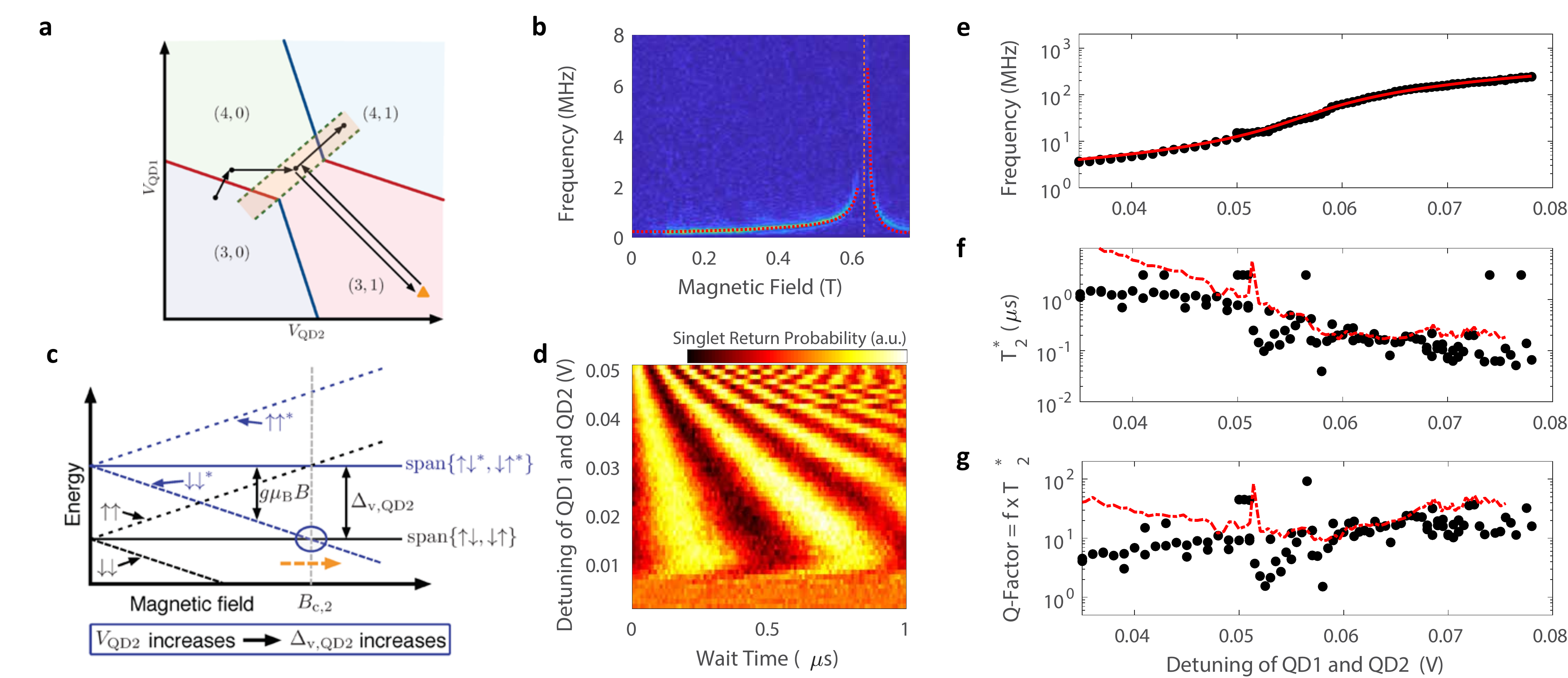}
	\caption{Spin-valley driven singlet-triplet rotations. (a) Schematic of the pulse sequence used to to interrogate the magnetic field and detuning voltage dependences of intervalley spin-orbit driven singlet-triplet rotations. (b) FFT of singlet-triplet rotations at fixed detuning in (3,1) as the external magnetic field is swept along the [010] crystallographic direction, with superimposed model fit (red dotted line). The orange dashed line indicates the spin-valley hot spot. We fit an intervalley SOC strength of $0.132 \pm 0.014 \ \mathrm{\mu eV}$ and a valley splitting of $73.177 \pm 0.033 \ \mathrm{\mu eV}$, with uncertainty reported here as 95\% confidence intervals (see Supplementary Information).
	(c) Magnetic field dependence of the system energy levels. The orange dashed arrow illustrates the change in the hot spot critical field as the QD-QD detuning is increased. 
	(d) Singlet-triplet rotations at a fixed magnetic field of 0.645 T as a function of QD-QD detuning voltage. (e) Measured singlet-triplet qubit rotation frequency, $f$, as a function of QD-QD detuning voltage (black circles), with superimposed model fit (red curve). We estimate a valley splitting lever arm of $46.25 \pm 0.85 \ \mathrm{\mu eV/V}$ (see Supplementary Information). (f) Black circles are extracted inhomogeneous dephasing times, $T_2^*$, as a function of QD-QD detuning. The red dashed curve is proportional to $|df/dV|^{-1}$, the expected dependence for quasi-static charge noise, where $|df/dV|$ is found from a numerical derivative of the data in (e).  (g) The black circles are the calculated quality factor of qubit rotations ($Q = f \times T_{2}^{*}$) as a function of QD-QD detuning. The red dashed curve is the expected quality factor from quasi-static charge noise found from the data in (e) and the red dashed curve in (f).}
	\label{fig:Exp:VoltageControl}
\end{figure*}

The system is initialized by loading a (4,0) singlet ground state, then quickly transferring an electron to the (3,1) charge configuration to produce a (3,1) singlet state (i.e. rapid adiabatic passage). Here, SOC in the DQD will drive rotations between (3,1) singlet and triplet states. We then rapidly return the system to the (4,0) charge sector, where Pauli spin blockade, combined with an  enhanced latching mechanism \cite{Harvey-Collard2018a}, is used to read out the spin state of the two-electron system in the single-triplet basis. In Fig.~\ref{fig:Exp:VoltageControl}(b) we show the fast Fourier transform (FFT) of SOC-driven rotations as the external magnetic field is swept along the [010] crystallographic direction. 
For low field strengths ($B < 0.5 \ \mathrm{T}$), we observe a weak increase in evolution frequency with applied magnetic field, consistent with previous experiments~\cite{Jock2018,Tanttu2019,Harvey-Collard2019}. In this regime an intravalley spin-orbit mechanism generates a difference in effective electron $g$-factor between the two QDs which lifts the degeneracy of the $m$=0 states, $\ket{\uparrow \downarrow}$ and $\ket{\downarrow \uparrow}$, driving rotations between (3,1)$S$ and (3,1)$T_0$~\cite{Jock2018}. 
As $B$ is further increased, we observed an unexpected rapid rise in rotation frequency with a sharp peak near $B$ = 0.64 T.  As discussed further below, these fast rotations are driven by an intervalley spin-orbit interaction which involves a coupling between distinct valley states having opposite spin. The peak position corresponds to the magnetic field at which the excited valley state, $T_{-}^{(1)}=\ket{\downarrow\downarrow^{(1)}}$, crosses the ground state $m$=0 manifold of the two-electron system, $B_{c,2}$, as illustrated in Fig. \ref{fig:Exp:VoltageControl}(c), often referred to as the spin-valley ``hot spot".  

Previous work has studied this regime in silicon QDs through single spin relaxation rates \cite{Yang2013,Hao2014,Borjans2019,Hollmann2019,Zhang2020}. 
Our approach of studying coherent rotations driven by the intervalley spin-orbit interaction yields new insight into the intervalley spin-orbit interaction and its dependence on applied magnetic field. Here we provide an intuitive three-level picture of the system that describes the physics of the frequency dependence around the hot spot in Fig.~\ref{fig:Exp:VoltageControl}(b) and explains how the system can be used for a novel qubit operating mode.  

In the two-electron DQD system, the intervalley hot spot corresponds to a distortion of the $m$=0 subspace $\lbrace \ket{\downarrow\uparrow}, \ket{\uparrow\downarrow} \rbrace$ of the ST qubit due to coupling to $\ket{\downarrow \downarrow^{(1)}}$, the down-polarized triplet state for which the electron in QD2 is in its excited eigenvalley.
This hybridizes the $\ket{\downarrow\uparrow}$ and $\ket{\downarrow \downarrow^{(1)}}$ states, while $\ket{\uparrow\downarrow}$ remains unperturbed. In the basis of $\lbrace \ket{\uparrow \downarrow}$, $\ket{\downarrow \uparrow}$, $\ket{\downarrow \downarrow^{(1)}} \rbrace$, this interaction can be represented by an effective three-level system with a Hamiltonian of the form 
\begin{equation}
\label{eq:HHotSpot}
H = \left(
\begin{array}{ccc}
B \delta & 0 & 0 \\
0 & -B \delta & \gamma \\
0 & \gamma^{*} & \Delta_{\mathrm{v,QD2}} - g_{*} \mu_{B} B
\end{array}
\right),
\end{equation}
where $\delta = \mu_{\mathrm{B}} \Delta g/2$, with $\Delta g = g_{1} - g_{2}$ the difference in $g$-factors between the QDs arising from variability of interfacial SOC, $\gamma$ is the intervalley coupling strength, $\Delta_{\mathrm{v,QD2}}$ is the valley splitting for the QD associated with the $\ket{\downarrow \downarrow^{(1)}}$ state, and $\mu_{\mathrm{B}}$ is the Bohr magneton. The $g$-factor governing the Zeeman shift of $\ket{\downarrow \downarrow^{(1)}}$ is $g_{*} = (g_{1} + g_{2}^{(1)})/2$, the average of the $g$-factors of the ground valley of QD1 and excited valley of QD2. The eigenstates of this three-level system are
\begin{eqnarray}
\ket{+} & = & w_{-} \ket{\downarrow\uparrow} + w_{+}\ket{\downarrow \downarrow^{(1)}} \\ 
\ket{\uparrow \downarrow} &  & \nonumber \\
\ket{-} & = & w_{+} \ket{\downarrow\uparrow} - w_{-}\ket{\downarrow \downarrow^{(1)}} \nonumber
\end{eqnarray}
where 
\begin{eqnarray}
w_{\pm} & = & \sqrt{1 \pm \eta/\sqrt{\eta^{2} + 4 \vert \gamma \vert^{2}}}/\sqrt{2} \\
\eta & = & \Delta_{\mathrm{v,QD2}} + (\delta \! - \! g_{*} \mu_{\mathrm{B}}) B,
\end{eqnarray}
with eigenenergies
\begin{eqnarray}
E_{\pm} & = & -B \delta + \frac{1}{2} \left(\eta \pm \sqrt{\eta^{2} + 4 \vert \gamma \vert^{2}} \right)\\
E_{\uparrow \downarrow} & = & B \delta \nonumber
\end{eqnarray}

The three-level Hamiltonian has three distinct energy gaps ($\Delta_{+} = E_{+} - E_{\uparrow \downarrow}$, $\Delta_{-} = E_{\uparrow \downarrow}-E_{-}$, $\Delta_{+-} = E_{+}-E_{-}$) and, in principle, three frequencies corresponding to the rate of dynamical phase accumulation for each of these gaps could be present in the measured spectrum. However, we observe only a single rotation frequency component in Fig.~\ref{fig:Exp:VoltageControl}(b). This can be understood by the following physical picture.
Supposing that the system is initially tuned away from the spin-valley anticrossing, the initial state prepared at the beginning of the evolution is close to $\ket{S} = \frac{1}{\sqrt{2}}(\ket{\uparrow \downarrow} - \ket{\downarrow \uparrow})$. If the valley splitting is changed to bring the system closer to the spin-valley hot spot, the $\ket{\downarrow \uparrow}$ state adiabatically deforms into either $\ket{+}$ or $\ket{-}$. 
The energy gap dictating the evolution frequency is the difference between $E_{\uparrow \downarrow}$ and the level (either $E_{+}$ or $E_{-}$) that is adiabatically connected to the initial $\ket{\downarrow \uparrow}$ state. When operating on the low-field (high-field) shoulder of the hot spot peak, the measured frequencies in Fig. \ref{fig:Exp:VoltageControl}(b) are dominated by rotations within the subspace spanned by $\lbrace \ket{{\uparrow \downarrow}}, \ket{{-}} \rbrace$ ($\lbrace \ket{{\uparrow \downarrow}}, \ket{{+}} \rbrace$), thus creating a two-level qubit system. Qubit measurement amounts to projecting back onto $\ket{S}$, with any support in the span of $\lbrace \ket{T_0}, \ket{\downarrow \downarrow^{(1)}} \rbrace$ read out as triplet.

\subsection{\textbf{Spin-valley driven singlet-triplet qubit}}

We realize this novel operating mode in the experiment by controlling the valley splitting of QD2 through modulation of the electric field \cite{Yang2013,Gamble2016,Hollmann2019,Zhang2020} at a constant magnetic field. We apply a field  of $B=0.645$ T along the [010] crystallographic direction, such that we are on the high magnetic field side of the hot spot peak ($g\mu_{\mathrm{B}}B > \Delta_{\mathrm{v}}$). 
In this case, an increase in applied electric field in QD2 will increase the valley splitting $\Delta_{\mathrm{v,QD2}}$, shifting the location of the spin-valley hot spot to higher magnetic field. We assume a linear dependence of valley splitting as a function of gate voltage away from a reference voltage $V_{0}$, $\Delta_{\mathrm{v,QD2}}(V) = \Delta_{\mathrm{v,QD2}}\vert_{V_{0}} + \lambda_{\mathrm{v}} (V_{\textrm{QD2}}-V_{0})$. We refer to $\lambda_{\mathrm{v}}$ as the valley splitting lever arm. Since we are operating at constant magnetic field, we would expect an increase in rotation frequency in the $\lbrace \ket{{\uparrow \downarrow}}, \ket{{+}} \rbrace$ subspace as we drive up the flank of the hot spot peak.

In Fig. \ref{fig:Exp:VoltageControl}(d) we show the singlet return signal as a function of time spent at the manipulation point in (3,1) as the QD-QD detuning, $\epsilon$, is varied along $\Delta V_{\mathrm{QD2}} = -\Delta V_{\mathrm{QD1}}$. The state is prepared in the same way as described above. For shallow detuning, we do not observe rotations since the exchange interaction, $J(\epsilon)$, is large and (3,1)S is nearly an eigenstate of the system. At moderate detuning we begin to see oscillations between singlet and triplet states due to the spin-orbit interaction, indicating a relative reduction in $J(\epsilon)$. As we pulse to deeper detuning, the voltage on the QD2 plunger increases. This enhances the vertical electric field confining QD2, resulting in an increase in valley splitting and hot spot critical field, $B_{\mathrm{c,2}} = \Delta_{\mathrm{v,QD2}} / g_{*} \mu_{\mathrm{B}}$, and an increase in rotation frequency (Fig. \ref{fig:Exp:VoltageControl}(d)). In Fig. \ref{fig:Exp:VoltageControl}(e) we plot the rotation frequency as a function of QD-QD detuning. Here, we demonstrate a rotation frequency in excess of 200 MHz, illustrating the ability to electrically control the intervalley spin-orbit driven frequency over a span of two orders of magnitude. Our model with an assumed linear dependence of valley splitting on gate voltage fits the data well, giving a valley splitting lever arm of $46.25 \pm 0.85 \ \mathrm{\mu eV/V}$. While large-scale implementation of this qubit approach will require some level of valley uniformity, valley splittings have been shown to be tunable by a few hundred $\mu$eV in MOS devices \cite{Yang2013,Gamble2016}, which eases this constraint and bolsters the prospects for future systems.

Next, we fit the decay in oscillations of measured singlet probability as a function of wait time, $t$, for a given detuning to a Gaussian envelope, $\exp(-(t/T_{2}^{*})^2)$. From this, we extract an inhomogeneous dephasing time, $T_{2}^{*}$, as a function of detuning, shown in Fig. \ref{fig:Exp:VoltageControl}(f). When the interaction with the excited valley is weak, we expect the dephasing to be dominated by the hyperfine interaction with 500ppm residual $^{29}$Si \cite{Assali2011,Witzel2012,Witzel2012b,Wu2014,Eng2015,Jock2018}. As the interaction strength increases, the coupling to nearby electric fields will be enhanced, increasing sensitivity to charge noise. For deeper detuning, we observe a decrease in $T_{2}^{*}$, which follows a $T_2^* \propto |df/dV|^{-1}$ dependence, depicted as a red dashed line in Fig. \ref{fig:Exp:VoltageControl}(f), which is expected for quasi-static charge noise \cite{Dial2013,Jock2018}. At frequencies above 100 MHz ($\epsilon >$ 65 mV), $T_2^*$ is lower than the expected fit for quasi-static charge noise. The spin-valley hot spot is known to lead to an enhanced spin relaxation rate \cite{Yang2013,Huang2014,Hao2014,Borjans2019,Hollmann2019,Zhang2020}, and may produce a T$_1$ limited dephasing as the system is tuned closer to the $S$-$T_-^{(1)}$ crossing.
We estimate a lower bound for such a $T_{1}$ time of no shorter 100 $\mathrm{ns}$, corresponding to a relaxation rate no faster than about 10 $\mathrm{MHz}$. This relaxation rate is orders of magnitude faster than measured hot spot $T_{1}$ times in the literature of around 1-200 $\mathrm{kHz}$ \cite{Yang2013,Veldhorst2014,Zhang2020}, but not inconsistent with extrapolated relaxation rates very close to the hot spot \cite{Zhang2020}.
The quality of rotations, $Q = f \times T_{2}^{*}$, which compares the rotation frequency to the dephasing time, is plotted in Fig. \ref{fig:Exp:VoltageControl}(g). We observe that, while the dephasing is faster at deeper detunings, the rotation frequency grows more quickly and the quality factor increases to $Q \sim 20$ at rotation frequencies above 100 MHz. We have observed hot spot driven rotation frequencies near 400 MHz in a separate $^{\textrm{nat}}$Si device, albeit with lower quality factors (see Supplementary Information). This highlights the dependence of the rotation quality on the interplay of the device tuning and the details of the intervalley coupling. Control of these parameters may provide a path to improving the rotation quality to produce higher-fidelity gate operations.

%% file: sec_Measurements/Measurements_HotSpotQubit.tex
\begin{figure*}[!]
	\centering
	\includegraphics[width=1.0\textwidth]{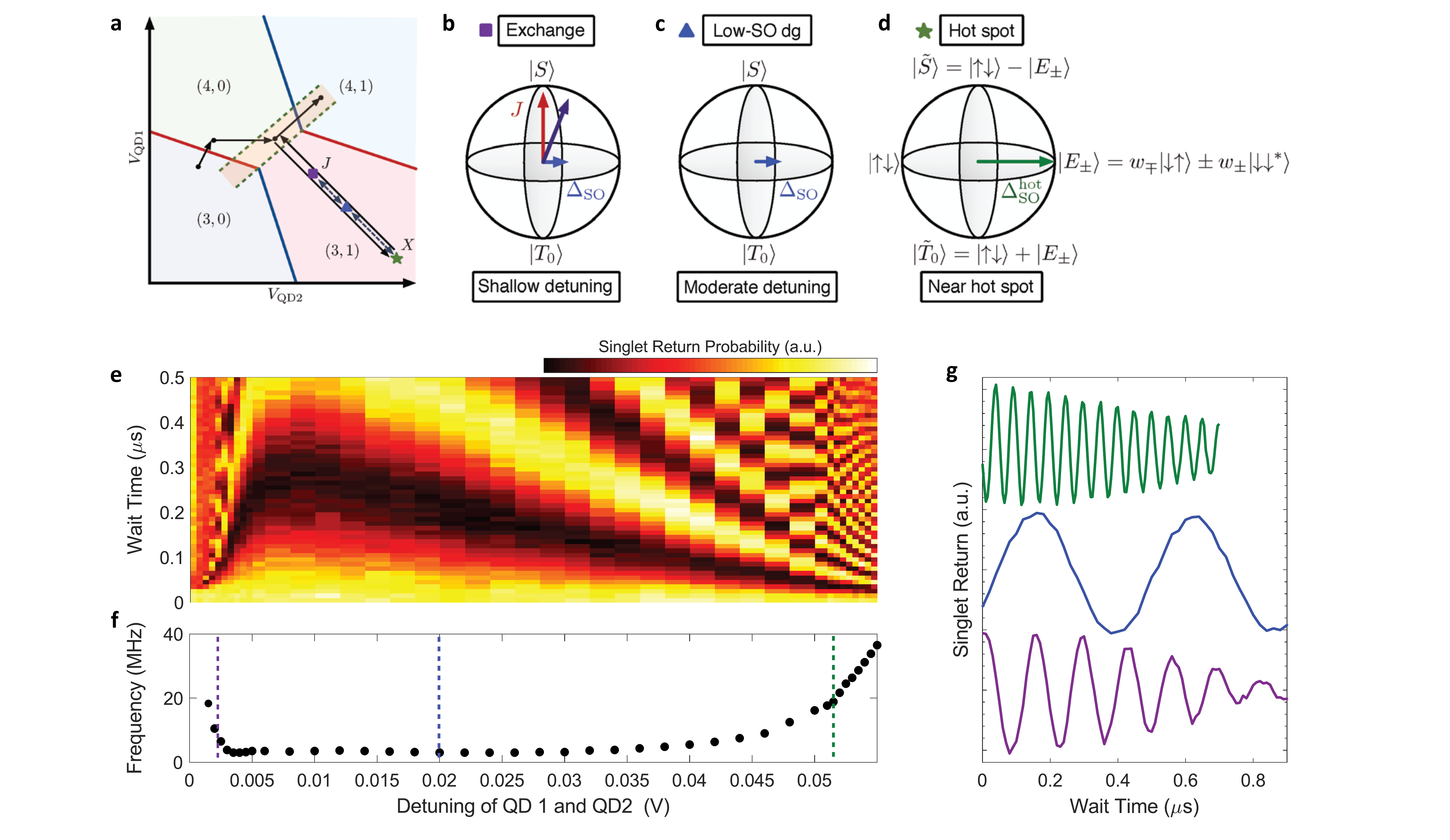}
	\caption{Two-axis singlet-triplet qubit control. (a) Schematic of the pulse sequence used to demonstrate all-electrical modulation between exchange-dominated and spin-orbit-dominated qubit control axes. Here rapid adiabatic passage is used to transfer the qubit to a (3,1)S state at moderate detuning, such that the strong spin-orbit effect is turned off, then the qubit is allowed to evolve for a $\pi/2$ rotation about the spin-orbit driven rotation axis. The qubit is then pulsed to some detuning point and allowed to evolve for some time, then returned to the original detuning and allowed to evolve for another $\pi/2$ rotation. The spin state of the qubit is then read out. The colored symbols represent three QD-QD detuning points of interest. (b-d) Bloch sphere representations of the qubit basis and control axes for the three QD-QD detuning points of interest. (e) Singlet return probability after performing qubit control rotations as a function of QD-QD detuning voltage at a fixed magnetic field of 0.645 T. (f) Extracted qubit rotation frequency from (e). (g) Qubit rotations for QD-QD detunings of 2.8 mV (purple), 20 mV (blue), and 55 mV (green), corresponding to exchange, low-SO, and hot spot driven qubit control, respectively.}
	\label{fig:Exp:TwoAxis}
\end{figure*}

The logical basis for singlet-triplet qubits is generally represented by the linear combination of $\ket{\uparrow \downarrow}$ and $\ket{\downarrow \uparrow}$ states (e.g., $\ket{S}$ and $\ket{T_0}$). During operation, the qubit states will rotate on the Bloch sphere about the vector sum of the Z-axis governed by the exchange energy, $J(\epsilon)$, and the X-axis dictated by the difference in Zeeman splitting between the two QDs, $\Delta E_{Z}$. Logic gates are performed by electrically pulsing between regions dominated by $J(\epsilon)$ and regions dominated by $\Delta E_{Z}$. In other implementations of singlet-triplet qubits, $\Delta E_{Z}$ is fixed \cite{Wu2014,Nichol2017,Harvey-Collard2017,Cerfontaine2019}. In contrast, by utilizing the electrically controlled intervalley interaction described above, we are able to independently implement high frequency spin-orbit driven gates at deep detuning, where the exchange interaction is weak, and exchange driven rotations at shallow detuning, where the intervalley interaction is weak and $J(\epsilon)$ dominates. 

In Fig. \ref{fig:Exp:TwoAxis} we demonstrate simultaneous two-axis control of the intervalley driven singlet-triplet qubit. We operate on the high-field shoulder of the spin-valley hot spot and define the qubit basis in terms of the $\ket{\uparrow \downarrow}$ and $\ket{{+}}$ states, where
\begin{eqnarray}
\ket{\tilde{S}} & = & \frac{1}{\sqrt{2}}(\ket{\uparrow \downarrow} - \ket{{+}}) \\
\ket{\tilde{T_0}} & = & \frac{1}{\sqrt{2}}(\ket{\uparrow \downarrow} + \ket{{+}}),
\end{eqnarray}
with $w_{+} \approx 0$ for shallow and moderate detuning, such that $\ket{\tilde{S}} \approx \ket{S}$ and $\ket{\tilde{T_0}} \approx \ket{T_0}$. Pulsing to shallow detuning drives exchange rotations, Fig. \ref{fig:Exp:TwoAxis}(b), while for deep detunings the intervalley spin-orbit interaction is turned on, Fig. \ref{fig:Exp:TwoAxis}(d). Furthermore, at moderate detuning (Fig. \ref{fig:Exp:TwoAxis}(c)), both the exchange and intervalley interactions are weak and spin interaction is dominated by the intravalley spin-orbit interaction \cite{Jock2018}. This provides a regime where qubit dephasing times are limited by the hyperfine interaction with residual $^{29}$Si in the host lattice and decoupled from charge noise. The ability to rapidly toggle between the two control axes by pulsing to detuning regions with large (small) exchange and small (large) spin-valley coupling, respectively, provides for high-orthogonality qubit control.

We can infer a qualitative measure of orthogonality of control over this qubit from the measurements shown in Figs. \ref{fig:Exp:TwoAxis}(e,f) and referring to an effective qubit Hamiltonian $H = h_{z} \sigma_{z} + h_{x} \sigma_{x}$. Since the exchange, $J(\epsilon)$, governs the $h_{z}$ component, while the intravalley and intervalley SOC control the $h_{x}$ component, as shown schematically in Figs. \ref{fig:Exp:TwoAxis}(b-d), the relative magnitudes of $h_{z}$ and  $h_{x}$ dictate the axis about which the qubit rotates on the Bloch sphere.
For moderate detuning (middle dashed line in Fig. \ref{fig:Exp:TwoAxis}(f)), where the intravalley SOC contribution, $\Delta_{\mathrm{SO}}$, dominates the $h_{x}$ component, we observe a qubit rotation frequency of $\sim$ 2 MHz. At high exchange operating point (near left dashed line in Fig. \ref{fig:Exp:TwoAxis}(f)), the qubit rotation frequency can reach $\sim$ 20 MHz. This corresponds to $h_{z} / h_{x} \approx 10$. Conversely, since the exchange $J(\epsilon)$ decays quickly with detuning $\epsilon$, 
the point of high intervalley SOC (right dashed line in Fig. \ref{fig:Exp:TwoAxis}(f)) corresponds to a region where the residual exchange is negligible. Here $h_{x}/h_{z} \gg 1$ and the axis of rotation is nearly on the equator of the Bloch sphere.

%% file: sec_Measurements/Measurements_ChargeNoise.tex
\subsection{\textbf{Characterization of MOS charge noise spectrum}}

\begin{figure}[t!]
	\centering
	\includegraphics[width=0.48\textwidth]{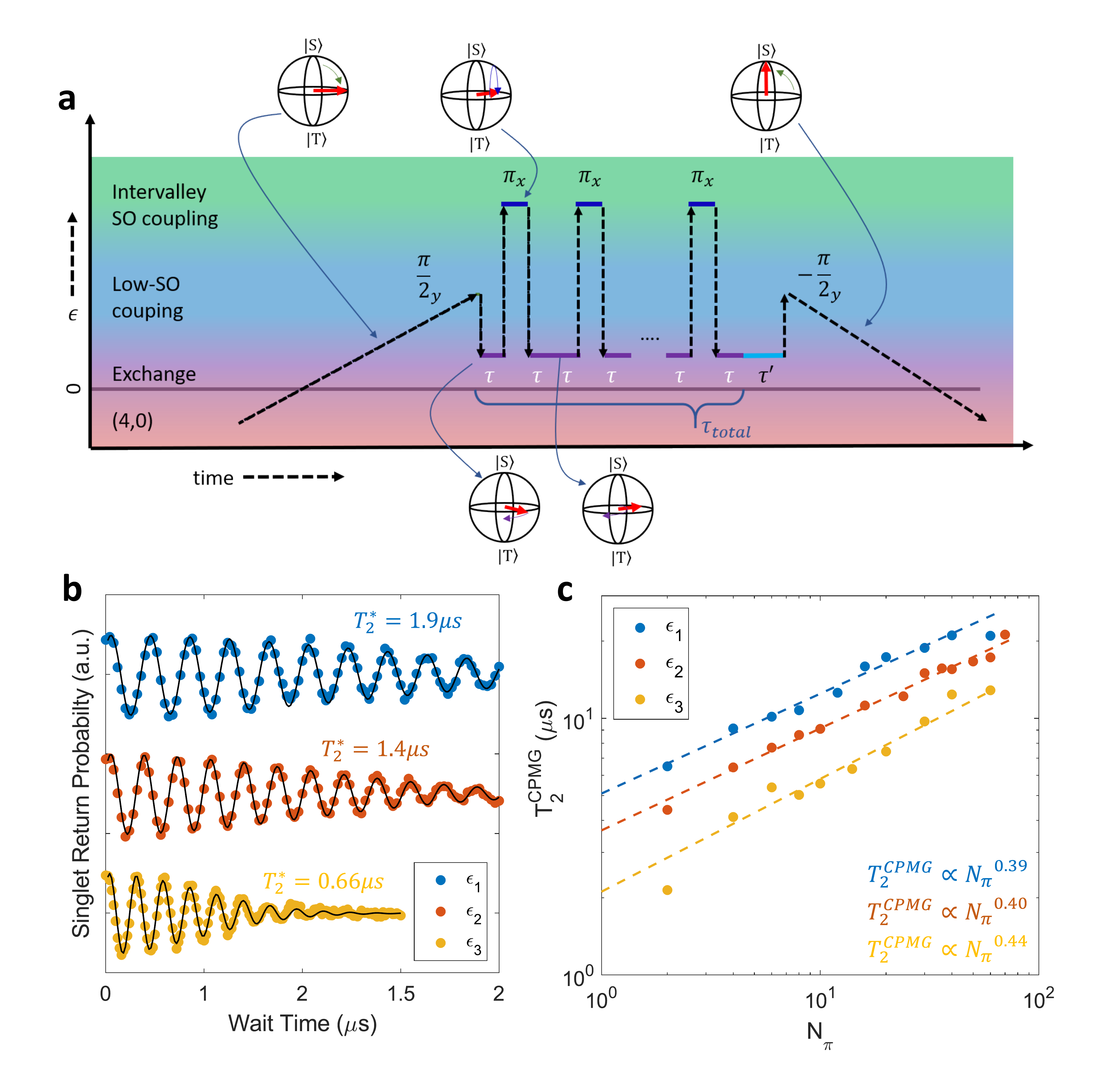}
	\caption{Decoupling from charge noise with CPMG. (a) Schematic for CPMG pulses. We initialize the qubit into the (4,0)S ground state and ramp adiabatically, such that the qubit transfers to the ground state ($\mathrm{\uparrow \downarrow}$ or $\mathrm{\downarrow \uparrow}$) in the (3,1) charge sector at moderate detuning, away from the spin-valley hot spot. This acts as an effective $\pi/2$ pulse about the Y-axis of the $S$-$T_0$ qubit basis. A fast pulse to and from a detuning, $\epsilon$, where exchange is substantial, drives coherent rotations around an axis dominated by the exchange interaction. Here, charge noise drives qubit dephasing. A series of $\pi$ pulses are then applied to decouple the qubit from charge noise. Here we operate with an intervalley spin-orbit driven rotation frequency of 20MHz. A final wait time, $\tau'$, at the end of the sequence allows for the observation of the free induction decay of the refocused echo. Returning to the (4,0) charge sector adiabatically produces an effective $-\pi/2$ Y-pulse and projects the states onto the (4,0)S and (3,1)T$_0$ basis for measurement. (b) Qubit exchange rotations at three QD-QD detuning points. Black curves are fits to oscillating Gaussian decay envelopes $\propto \exp\left(-(t/T_{2}^{*})^2\right)$. (c) Qubit CPMG coherence time as a function of the number of refocusing pulses $N_{\pi}$ for three QD-QD detuning points where exchange is the dominant spin interaction. Dashed lines are fits to the form $T_{2}^{\textrm{CPMG}} \propto N_{\pi}^{\beta}$.}
	\label{fig:Exp:CPMGEcho}
\end{figure}

Having demonstrated high-orthogonality all-electrical control over fast Z (exchange) and X (spin-orbit) gates, we turn now to exploiting these fast  operations to probe the spectral content of noise in our device at relatively high frequencies. In silicon QD based spin qubits, where magnetic noise may be reduced by the use of enriched $^{28}$Si, charge noise has been identified as a dominant source of error \cite{Beaudoin2015}. 
Here, charge noise may have the effect of increasing dephasing rates for one- or two-qubit gates involving the exchange interaction or when the architecture employs a magnetic field gradient from a micromagnet for spin control. 
CPMG pulse sequences are a well-established technique for mitigating the effects of qubit dephasing by applying a series of refocusing control pulses \cite{Carr1954,Meiboom1958} and has been successfully demonstrated with silicon spin qubits \cite{Medford2012,Muhonen2014,Chan2018,Yoneda2018,Connors2021}. In Fig. \ref{fig:Exp:CPMGEcho} we demonstrate the ability to use a CPMG pulse sequence to prolong the qubit coherence time. We apply a string of intervalley spin-orbit driven pulses to decouple the qubit from charge noise during the spin-spin exchange interaction. Fig. \ref{fig:Exp:CPMGEcho}(b) shows qubit exchange rotations for three QD-QD detuning voltages. We see that for faster exchange pulses the qubit dephases more quickly, as expected for quasistatic charge noise dominated inhomogeneous dephasing in qubit exchange gates \cite{Dial2013,Petersson2010,Shi2013,Thorgrimsson2017,Jock2018}. Fig. \ref{fig:Exp:CPMGEcho}(c) shows the CPMG coherence time, $T_{2}^{\textrm{CPMG}}$, versus the number of refocusing pulses, $N_{\pi}$, for the three detuning values. We observe an increase in coherence time with increasing $N_{\pi}$, which follows a power-law dependence with $T_{2}^{\textrm{CPMG}} \propto N_{\pi}^{\beta}$. We find exponents of $\beta \approx 0.39, 0.40$, and $0.41$ for the detuning points $\epsilon_1$, $\epsilon_2$, and  $\epsilon_3$, respectively. CPMG prolongs qubit coherence by refocusing noise for time scales longer than the time between refocusing pulses. For a given total time exposed to exchange, $\tau_{\textrm{total}}$, more $N_{\pi}$ pulses will decrease the time the qubit is exposed to noise before being refocused. As such, the effectiveness of CPMG to mitigate charge noise will largely be determined by the noise spectral density, $S(f)$. For colored noise of the form $S(f) \propto f^{-\alpha}$, we expect $T_{2}^{\textrm{CPMG}} \propto N_{\pi}^{\frac{\alpha}{1+\alpha}}$ \cite{Medford2012,Muhonen2014,Chan2018,Yoneda2018,Jirovec2021,Connors2021}. A fit to the data in Fig. \ref{fig:Exp:CPMGEcho}(c) indicates a noise spectrum with $\alpha \approx$ 0.7.

\begin{figure*}
	\centering
	\includegraphics[width=0.99\textwidth]{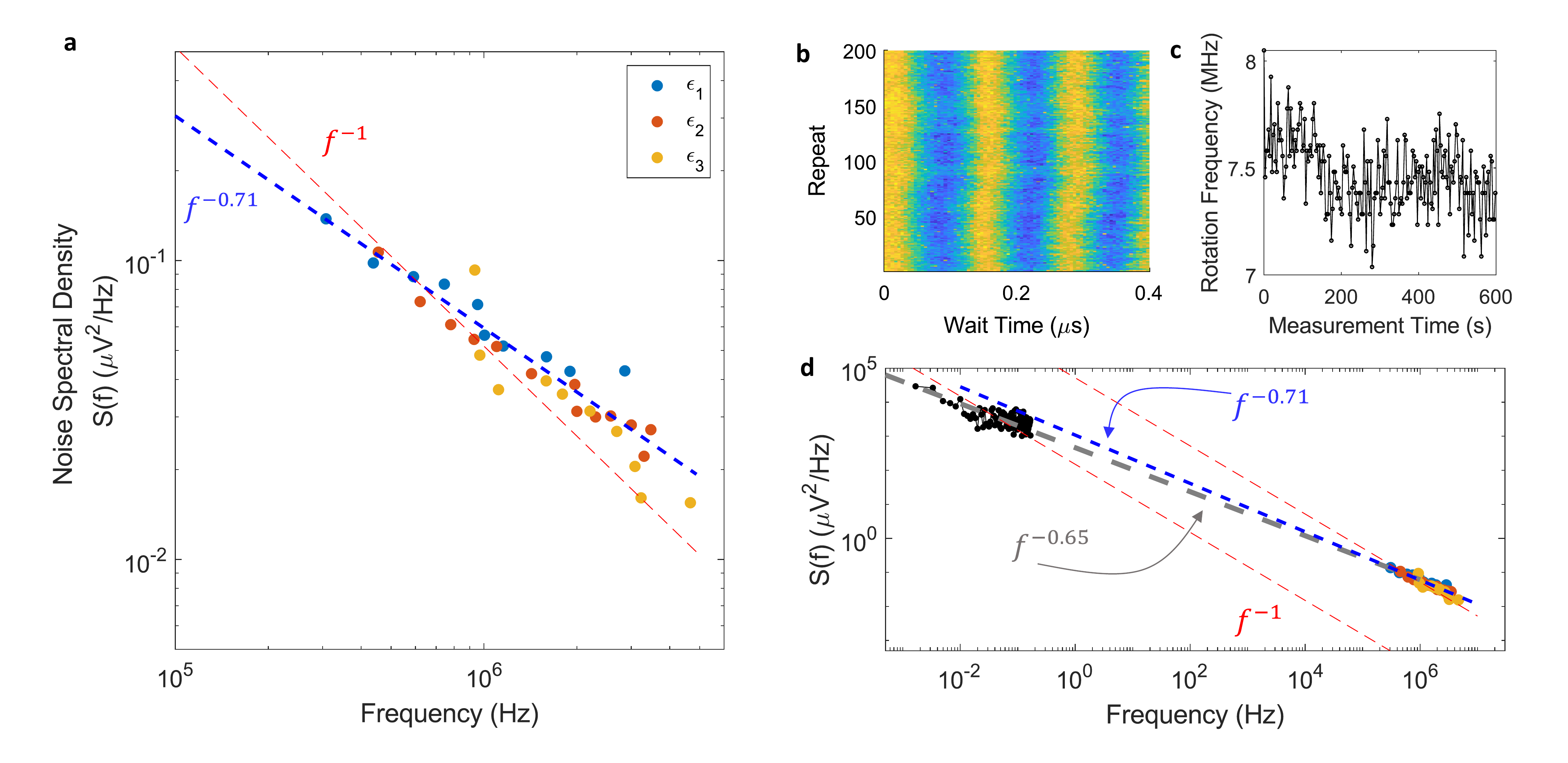}
	\caption{Charge noise spectrum. (a) Noise spectral density for charge noise experienced by the qubit during exchange pulses for three QD-QD detuning values. The blue dashed line is a power law fit to the data, $S(f) \propto f^{-\alpha}$. The red dashed line is a fit to a $1/f$ noise spectrum as a guide to the eye. (b) Repeated experiment of singlet return probability versus wait time for an exchange pulse near detuning $\epsilon_3$ over the course of 10 minutes. (c) Extracted qubit frequency for data in (b) as a function of experimental measurement time. (d) combined low- and high-frequency measurements of the noise spectral density. The blue dashed line is a power law fit to the high frequency data and the gray dashed line is a power law fit to the low frequency data extracted from (c). The red dashed lines are fits of $1/f$ spectra to the low- and high-frequency data sets, respectively, as guides to the eye.}
	\label{fig:Exp:PSD}
\end{figure*}

Treating the CPMG sequence as a noise filter \cite{Biercuk2011,Bylander2011,Alvarez2011} provides a noise spectroscopy method to determine the noise power spectral density (PSD). This technique has been utilized to characterize other solid-state qubits \cite{Bylander2011,Muhonen2014,Chan2018,Yoneda2018,Connors2021}. Considering the first harmonic of a bandpass filter, the strength of the noise PSD, for a given data point in Fig. \ref{fig:Exp:CPMGEcho}(c), is given by 
\begin{equation}
S(f_{N_{\pi}}) = \frac{\pi^2}{4 \cdot T^{\textrm{CPMG}}_{2,N_{\pi}}},
\label{eq:nPSD_Power}
\end{equation}
where $f_{N_{\pi}}$ is the relevant noise frequency being interrogated and is given by the time between pulses when refocused echo intensity drops to $1/e$,
\begin{equation}
f_{N_{\pi}} = \frac{N_{\pi}}{T^{\textrm{CPMG}}_{2,N_{\pi}}}.
\label{eq:nPSD_freq}
\end{equation}
The noise PSD is given in terms of fluctuations in exchange rotation frequency, which will be dependent on the strength of the exchange interaction at each detuning value. By using the gradient of the qubit frequency at each detuning point, $df(\epsilon)/dV(\epsilon)$, we convert the spectrum to voltage noise on the QD-QD detuning, which provides a means to compare the three detuning points. The combined data are plotted in Fig. \ref{fig:Exp:PSD}(a), where a strong agreement in the noise PSD for all three detuning values is observed. The blue dashed line is a power law fit, which gives $S(f) \propto f^{-0.71}$.

Next, we examine the low-frequency portion of charge noise spectrum in this system. In Fig. \ref{fig:Exp:PSD}(b) we plot the singlet return probability for repeated exchange rotation experiments near detuning $\epsilon_3$. Fig. \ref{fig:Exp:PSD}(c) shows the slow drift in the extracted exchange rotation frequency. Using a periodogram method and $df(\epsilon)/dV(\epsilon)$ at this tuning, we plot the low frequency noise PSD in Fig. \ref{fig:Exp:PSD}(d) alongside the high frequency results. A power law fit to the low frequency data (gray dashed line), extracted out to high frequency shows a $S(f) \propto f^{-\alpha}$ dependence of the charge noise PSD with $\alpha \approx 0.7$ in the mHz to MHz frequency range. This is consistent with what is observed for single QDs in semiconductors, where the charge noise is often found to be $1/f$-like, with $\alpha$ near 1 \cite{Nadj2010b,Chan2018,Kawakami2016,Freeman2016,Yoneda2018,Kim2019,Rudolph2019,Struck2019,Connors2019,Kranz2020,Connors2021,Hendrickx2021} and presumed to be caused by a distribution of charge fluctuators. We perform analogous measurements to characterize the power spectral density of magnetic noise and find a $S(f) \propto f^{-1.66}$ power law dependence\cite{Eng2015,Struck2019} (see Supplementary Information).

%% file: sec_Discussion/Discussion.tex
Interfacial spin-orbit interactions are known to play a significant role in the control of spin qubits in silicon QDs. In this work, we observe a rapid increase in the singlet-triplet rotation frequency near the spin-valley hot spot and develop a simple three state model to explain the observations. We utilize this effect to demonstrate an intervalley driven singlet-triplet qubit with high-orthogonality and fast electrical-only qubit control. We show the ability to electrically tune the intervalley spin-orbit interaction, enabling high-speed modulation between three qubit control regimes: (1) large exchange interaction, (2) small effective magnetic field gradient between QDs, and (3) hot spot driven qubit rotations with operational rotation frequencies exceeding 200 MHz. When the intervalley spin-orbit or exchange interactions are weak, qubit dephasing is dominated by the hyperfine interaction with the residual $^{29}$Si in the isotopically enriched substrate. However, for strong exchange or intervalley spin-orbit coupling, quasi-static charge noise becomes the dominant dephasing mechanism. This is the first experimental demonstration utilizing control of the spin-valley coupling to coherently drive a silicon spin qubit, establishing the intervalley driven singlet-triplet qubit as a candidate for future quantum information processing platforms.

Additionally, we highlight the utility of this novel qubit operating mode to probe specific physical phenomena relevant to silicon-based qubit platforms. Fits to our three-state model allow for an extraction of a valley splitting of $73.177 \pm 0.033 \ \mathrm{\mu eV}$ with a valley splitting lever arm of $46.25 \pm 0.85 \ \mathrm{\mu eV/V}$ and an intervalley SOC strength of $0.132 \pm 0.014 \ \mathrm{\mu eV}$. Furthermore, we exploit the filter function properties of CPMG dynamical decoupling techniques to extract the noise power spectrum of the charge noise in this device without the added complexity of on-chip, nano-fabricated micromagnets or nearby co-planar striplines that may otherwise be needed for qubit control. The fast hot spot refocusing pulses and strong coupling to charge noise when the exchange interaction is turned on allows for a probe of the noise power spectral density at high frequencies. These experiments, combined with low frequency drift measurements, reveal a noise spectrum  consistent with $S(f) \propto f^{-0.7}$ for frequencies between 3 mHz and 3 MHz.

%% file: sec_Methods/Methods.tex
\subsection{Device overview}
The double quantum dot studied in this work was realized in a fully foundry-compatible, single-gate-layer, silicon metal-oxide-semiconductor (MOS) device structure containing an epitaxially-enriched $^{28}$Si layer with 500 ppm residual $^{29}$Si at the Si/SiO$_2$ interface. The confinement and depletion gates are defined by electron beam lithography followed by selective dry etching of the poly-silicon gate layer, which produces the pattern shown in  Fig. \ref{fig:Device}(a). Electrons are confined at the Si/SiO$_2$ interface and relevant biasing of the poly-silicon gates create quantum dot potentials under the tips of gates QD1 and QD2. The tunnel rate to the electron reservoirs under the large gates in the bottom left and top right corners of the device is controlled by the applied voltage to the reservoir gates \cite{Rochette2019}. We operate with the bottom left electron reservoir receded such that the DQD system is coupled only to the top right reservoir through QD1. A single electron transistor (SET) in the lower right corner of the device is used for charge sensing. The number of electrons in each QD is inferred from changes in current through the SET as well as by magneto- and pulsed-spectroscopy methods.
\subsection{Measurements}
Measurements were performed in a $^3$He/$^4$He dilution refrigerator with a base temperature of around 8 mK. The effective electron temperature in the device was 150 mK. Gates QD1 and QD2 are connected to cryogenic RC bias-T's, which allow for the application of combined DC bias voltages and fast gate pulses. An external magnetic field is applied using a 3-axis vector magnet. We perform cryogenic preamplification of the charge sensing SET current using a heterojunction bipolar transistor (HBT) \cite{Curry2019}.

%% file: sec_Appendix/Appendix.tex
\begin{figure*}[h!]
	\centering
	\includegraphics[width=0.95\textwidth]{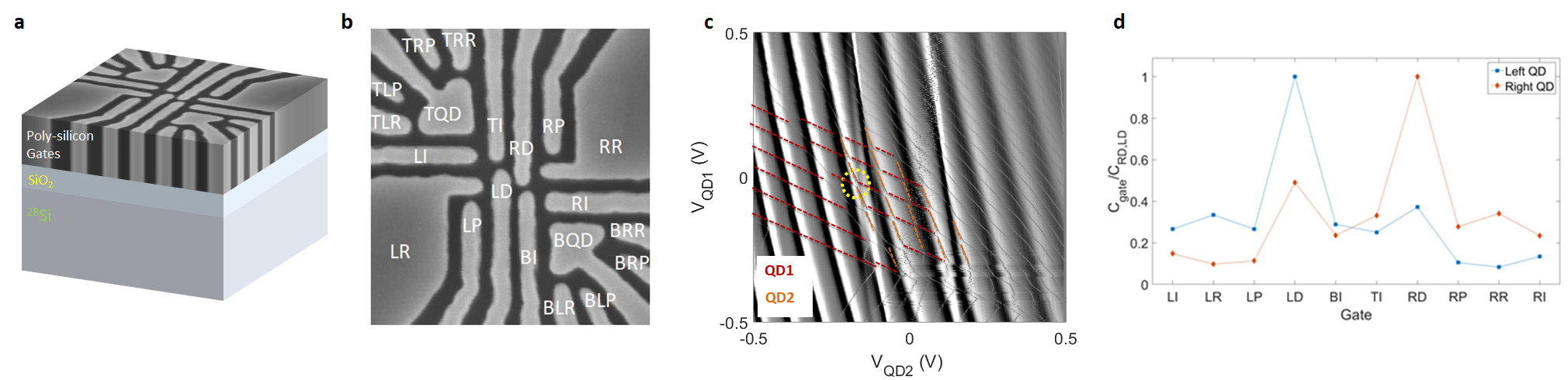}
	\caption{MOS DQD device structure. (a) A cartoon schematic of the MOS gate stack. (b) A top-down SEM of the single-layer poly-silicon gate design with the gate names labeled in white.(c) Charge stability diagram of the DQD. Here, we plot the gradient of the SET charge sensor current as the gates QD1 and QDs are varied about fixed offset voltages (V$_{\textrm{DC,QD1}}$ = 3.4 V and V$_{\textrm{DC,QD2}}$ = 3.9 V). The broad diagonal background features are due to Coulomb blockade peaks of the SET charge sensor. The sharp features correspond to charge transitions in the QDs. The red and orange dashed lines are guides to the eye for QD$_1$ and QD$_2$ charge transitions. The yellow circle represents the (N$_{\textrm{QD1}}$,N$_{\textrm{QD2}}$) = (4,0)-(3,1) charge region where this work was done. (d) Gate capacitance to QDs relative to LD (blue) and RD (red) gates.}
	\label{fig:Appendix:Device}
\end{figure*}

\subsection{Measurement details}
\label{sec:Appendix:Measurement}

The DQD studied in this work was formed within a device nominally identical to that shown in Supplementary Fig. \ref{fig:Appendix:Device}(a,b). This device was fabricated in a fully foundry-compatible process using a single-gate-layer, metal-oxide-semiconductor (MOS) poly-silicon gate stack with an epitaxially-enriched $^{28}$Si epi-layer with 500 ppm residual $^{29}$Si. The device is operated in enhancement mode using voltage biasing of the highly doped n+ poly-silicon gates to confine electrons to quantum dot (QD) potentials under gates RD and LD (QD1 and QD2, respectively). The gates LR, RR, BLR, BRR, TLR, and TRR overlap with implanted n+ ohmic contacts regions and are biased to accumulate two-dimensional electron gas (2DEG) regions under each gate that serve as electron reservoirs for the quantum dots and single electron transistor (SET) charge sensors. The upper left and bottom right corners of the device may be used as charge-sensing SETs by confining QDs under gates TQD and BQD, respectively. In the measurements presented in the main text, we utilize only the bottom right SET.

The number of electrons in each QD may be inferred from changes in current through the SET, as depicted in Supplementary Fig. \ref{fig:Appendix:Device}(c). The collection of red and orange parallel lines correspond to charge transitions in QD1 and QD2, respectively. We can infer the approximate locations of QD1 and QD2 by measuring their capacitances to nearby poly-silicon gates through scans of voltages applied to pairs of poly-silicon gate electrodes. From these measurements, we obtain the relative capacitance of the QDs to each gate compared to their capacitance to RD or LD, the gates that have strongest capacitive coupling to QD1 and QD2, respectively. We plot the relative capacitances in Supplementary Fig. \ref{fig:Appendix:Device}(d). The measured symmetric capacitance ratios of the two QDs indicate that they are well-formed lithographic quantum dots.

\begin{figure*}[!]
	\centering
	\includegraphics[width=0.95\textwidth]{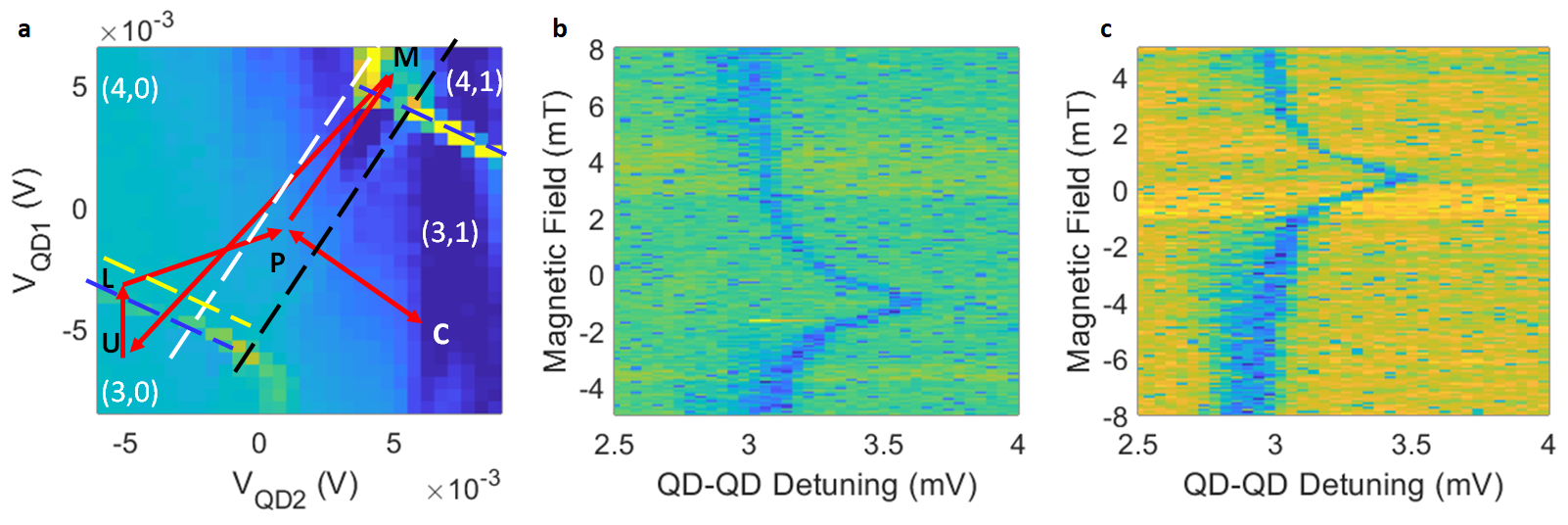}
	\caption{Device operation and spin funnels. (a) A pulsed charge stability diagram for the (4,0)-(3,1) anticrossing, showing the gradient of the charge sensor current. The red arrows depict a general pulse sequence for controlling the qubit: The system is initialized by first unloading an electron from the DQD (point U). An energy-selective pulse is applied to load a (4,0)S ground state (point L). The system is then plunged (point P) near the charge anti-crossing. The electrons are then separated (point C) and qubit manipulation pulse sequences are performed in the (3,1) charge region. The system is then pulsed back to point P where, due to Pauli spin blockade, a singlet spin state is allowed to transfer to the (4,0) charge state but a triplet spin state is energetically blocked and remains in a (3,1) charge state. An enhanced latching mechanism is then utilized for a spin-to-charge conversion (point M). Here the qubit control point, C, may consist of a complex voltage detuning sequence for qubit manipulation. The black and white dashed lines correspond to the location of the singlet and triplet state inter-QD charge preserving lines, respectively. (b,c) Spin funnel measurements indicating degeneracy between (3,1)S and (3,1)T$_-$ states as a function of QD-QD detuning voltage for magnetic fields applied along the [100] and [110] in-plane crystallographic directions, respectively.}
	\label{fig:Appendix:PulseLevels}
\end{figure*}


We operate this system near the (N$_{\textrm{QD1}}$,N$_{\textrm{QD2}}$) = (4,0)-(3,1), spin-blockaded  charge anti-crossing. A charge stability diagram for the double dot system is shown in Supplementary Fig. \ref{fig:Appendix:Device}(c). The ground state charge configuration is determined by the detuning between dots, $\epsilon$, which is controlled by tuning the voltages on gates RD and LD. These gates are connected to cryogenic RC bias-T’s which enable the application of fast gate pulses. A schematic of the cyclical pulse sequence used in qubit measurements is shown in Supplementary Fig. \ref{fig:Appendix:PulseLevels}(a). The system is initialized in the (4,0) charge sector by first unloading an electron from the DQD (point U). An energy-selective pulse into the (4,0) charge state between the singlet and triplet energy levels is applied to load a (4,0)S ground state (point L). Following that, the system is plunged (point P) to a detuning ($\epsilon$ $<$  0) close to the charge anti-crossing. The electrons are then separated (point C) and qubit manipulation pulse sequences are performed in the (3,1) charge region ($\epsilon$ $>$ 0). The system is then pulsed back to the (4,0) charge sector (point P) where, due to Pauli spin blockade, a singlet spin state is allowed to transfer to the (4,0) charge state but a triplet spin state is energetically blocked and remains in a (3,1) charge state \cite{Petta2005}. We then use an enhanced latching mechanism for a spin-to-charge conversion (pulsing to point M)\cite{Harvey-Collard2018a,Petersson2010,Studenikin2012,Mason2015,Nakajima2017,Broome2017,Harvey-Collard2017}. This technique relies on two tunneling events to load an electron onto QD2 from the reservoir, since an electron must first tunnel through QD1. This causes singlet states to remain locked in a metastable (4,0) charge state at point M, as a slow co-tunneling process is required to equilibrate. On the other hand, triplet states may quickly transfer to (4,1) by inelastic tunneling of an electron from the lead onto QD1. 

\subsection{CPMG analysis}

The measured data for a CPMG echo experiment at detuning $\epsilon_3$  with $N_{\pi} = 10$ are shown in Supplementary Fig. \ref{fig:Appendix:EchoToPSD}(a). Here, we plot the singlet return probability for time $\tau'$ after the end of the CPMG sequence as the total qubit evolution time $\tau_{\textrm{total}}$ increases. The oscillations in singlet return represent the free induction decay, FID, of the refocused echo. Fitting the FID to a Gaussian envelope function for each $\tau_{\textrm{total}}$ gives the echo amplitude as a function of total time exposed to charge noise. Supplementary Fig. \ref{fig:Appendix:EchoToPSD}(b) plots this as the number of refocusing pulses, $N_{\pi}$, is increased. As described in the main text, the relevant noise frequency being interrogated is $f_{N_{\pi}}$. The coherence time  $T^{\textrm{CPMG}}_{2,N_{\pi}}$, when the echo drops to $1/e$, indicates the noise strength near that frequency \cite{Bylander2011,Muhonen2014,Chan2018,Yoneda2018}. In Supplementary Fig. \ref{fig:Appendix:EchoToPSD}(c) the noise power spectral density is plotted for the three detuning values shown in the main text.

\begin{figure*}[!]
	\centering
	\includegraphics[width=0.9\textwidth]{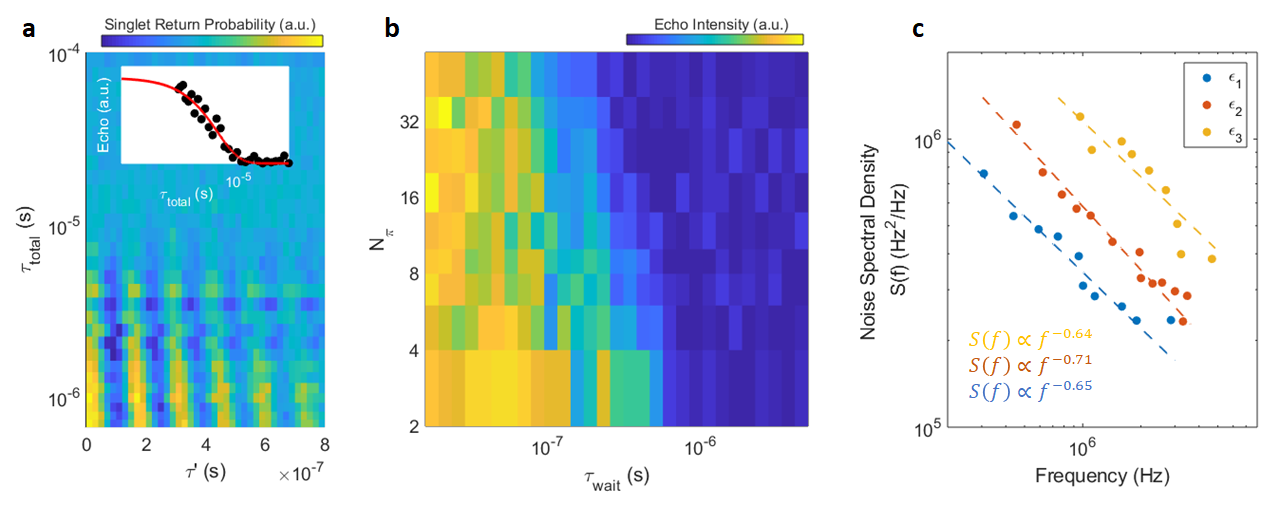}
	\caption{CPMG experiments. (a) CPMG echo at detuning $\epsilon_3$  with $N_{\pi} = 10$. We plot singlet return probability as function of wait time after the qubit is refocused, $\tau'$, as a function of $\tau_{\textrm{total}}$, the total qubit manipulation time for a CPMG sequence. For each $\tau_{\textrm{total}}$, the echo can be fit to an oscillating Gaussian decay to extract the echo amplitude. (a,inset) Echo amplitude as a function of total time exposed to charge noise, $\tau_{\textrm{total}}$,  for $N_{\pi} = 10$. Red line is a fit to a decay of the form $\exp(-(t/T_{2}^{*})^n)$. (b) Echo amplitude as a function of wait time, $\tau_{\textrm{wait}}$, as $N_{\pi}$ is stepped. (c) Frequency noise spectral density for charge noise experienced by the qubit during exchange pulses for three QD-QD detuning values. The dashed lines are power law fits to the data.}
	\label{fig:Appendix:EchoToPSD}
\end{figure*}

\begin{figure*}[!]
	\centering
	\includegraphics[width=0.9\textwidth]{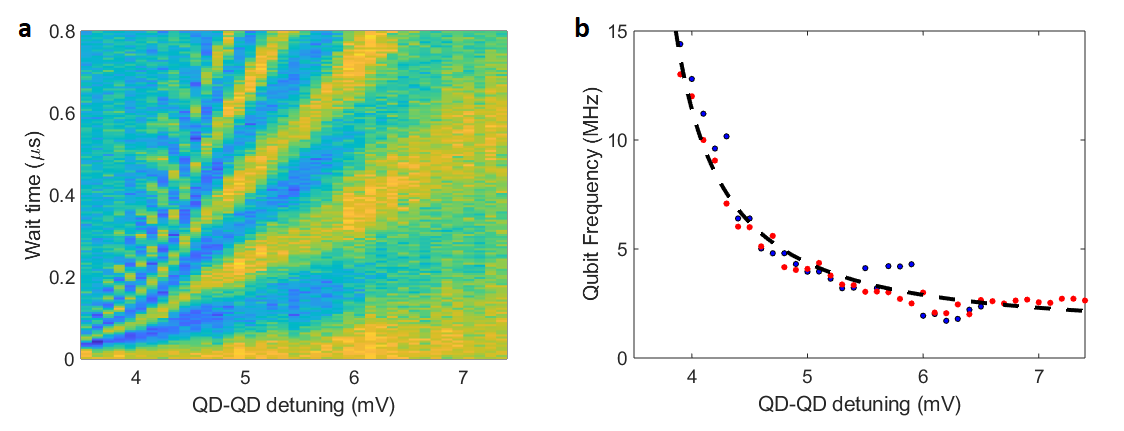}
	\caption{DQD exchange. (a) Exchange rotations at a fixed magnetic field of 0.645 T as a function of QD-QD detuning voltage. (b) Extracted frequency of qubit exchange rotations as a function of QD-QD detuning voltage. The blue and red circles are two experimental data sets and the black dashed line is a fit to the form $f(\epsilon) = \sqrt{J(\epsilon)^2 + \Delta E_Z^2}$, with $J(\epsilon) \propto \frac{t_c^2}{\epsilon}$.}
	\label{fig:Appendix:FvEps}
\end{figure*}

Supplementary Fig. \ref{fig:Appendix:FvEps} shows exchange dominated ST qubit rotations, reflected in the singlet return probability as a function of time the exchange interaction is turned on for a range of QD-QD detunings. Here, we initialize the qubit into the (4,0)S ground state and ramp adiabatically into the (3,1) charge region, such that it transfers to the ground state,  $\ket{\uparrow\downarrow}$ or $\ket{\downarrow\uparrow}$. A rapid pulse to and from a detuning, $\epsilon$, where exchange is substantial drives coherent rotations around an axis depending on both exchange, $J(\epsilon)$, and the difference in Zeeman splitting, $\Delta E_Z$. Returning to the (4, 0) charge sector adiabatically projects the states onto the (4,0)S and (3,1)T$_0$ basis for measurement. We then fit the rotation frequency, $f$, at each detuning to a smooth function to find the derivative, $df/dV$. This is used to convert the noise power spectral density in Supplementary Fig. \ref{fig:Appendix:EchoToPSD}(c) from a frequency to a voltage fluctuation, which allows for a comparison of noise power at the measured detuning points shown in the main text. 

\subsection{Magnetic noise}
\label{sec:Appendix:MagneticNoise}

\begin{figure*}[!]
	\centering
	\includegraphics[width=0.9\textwidth]{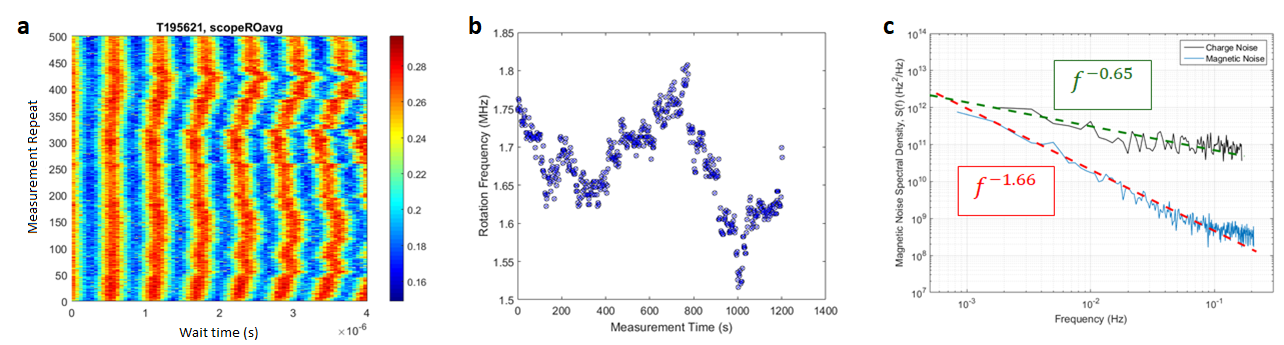}
	\caption{Low frequency magnetic noise. (a) Repeated experiment of singlet return probability versus wait time for spin-orbit driven singlet-triplet rotations over the course of 20 minutes.  (b) Extracted qubit frequency for data in (a) as a function of experimental measurement time. (c) The low frequency noise spectrum extracted using a periodogram method for magnetic (blue data) and charge (black data) noise. The red and green dashed lines are fits to $S(f) \propto f^{-\alpha}$ noise spectra.}
	\label{fig:Appendix:MagneticNoise1}
\end{figure*}

We used similar techniques to characterize the power spectral density of magnetic noise in Device A. Here we look at singlet-triplet rotations at shallow detuning away from the hot spot. In Supplementary Figs. \ref{fig:Appendix:MagneticNoise1}(b,c) we show singlet return probability for repeated experiments of SOC-driven qubit rotations and their extracted rotation frequency over the course of 20 minutes. In Supplementary Fig. \ref{fig:Appendix:MagneticNoise1}(c) we plot the noise PSD, which displays a $S(f) \propto f^{-1.66}$ power law dependence \cite{Eng2015,Struck2019}.

Next we utilize a CPMG sequence to decouple the qubit from  magnetic noise and examine the PSD at higher frequencies. To do this, we use an exchange $\pi$ pulse at a qubit frequency of 8.33 MHz, illustrated in Supplementary Fig. \ref{fig:Appendix:MagneticNoise2}. When the number of refocusing pulses, $N_{\pi}$, is increased, we observe no change in the CPMG coherence time, $T_{2}^{\textrm{CPMG}}$. This suggests a flat (white) noise mechanism, which is consistent with the extracted noise power spectral density. In this low-frequency regime coherence may be limited by a T$_1$ qubit relaxation process, but to be definitive further studies are needed.

\begin{figure*}[!]
	\centering
	\includegraphics[width=0.95\textwidth]{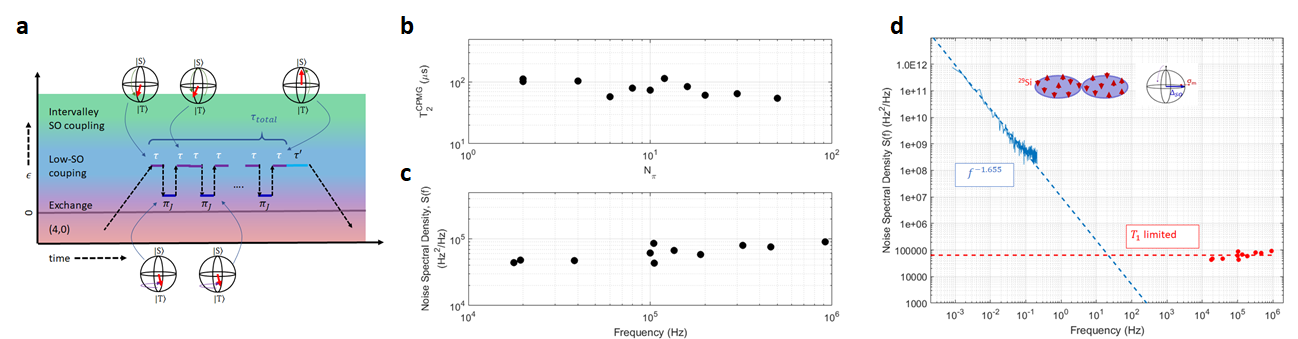}
	\caption{Decoupling from magnetic noise with CPMG. (a) Schematic for CPMG pulses to investigate magnetic noise. We initialize the qubit into the (4,0)S ground state and transfer one electron to the neighboring dot using rapid adiabatic passage, such that the qubit remains a singlet in the (3,1) charge sector. The qubit is then allowed to evolve and dephase due to fluctuations in the Overhauser fields between the two QDs. A series of fast pulses to and from a detuning $\epsilon$, where J is substantial, drive $\pi/2$ pulses to decouple the qubit from magnetic noise. A final wait time, $\tau'$ at the end of the sequence allows for the observation of the free induction decay of the refocused echo. Returning to the (4,0) charge sector by rapid adiabatic passage projects the states onto the (4,0)S and (3,1)T$_0$ basis for measurement. (b) Qubit CPMG coherence time as a function of the number of refocusing pulses, $N_{\pi}$. (c) High frequency noise spectrum for magnetic noise experienced by the qubit. (d) Combined low- and high-frequency measurements of the magnetic noise power spectral density. The blue and red dashed lines are fits to a power law, $S(f) \propto f^{-1.66}$, and constant (white noise), respectively. (inset) Fluctuations in the nuclear spins of the residual $^{29}$Si in each QD will cause fluctuations in their respective Zeeman splittings and cause dephasing of singlet-triplet qubit rotations.}
	\label{fig:Appendix:MagneticNoise2}
\end{figure*}

\subsection{Supporting measurements}
\label{sec:Appendix:Martin}
We independently observed valley hot spot-driven singlet-triplet rotations in another silicon device, Device B, shown in Supplementary Fig. \ref{fig:Appendix:Martin1}. This device was fabricated similarly to the device presented in the main text, but with two main differences: (1) This device has a natural silicon substrate and (2) uses a single accumulation gate SET charge sensor design \cite{Tracy2016,Harvey-Collard2017,Harvey-Collard2018a} for its bottom right charge sensor. The device was operated in a similar fashion near the (N$_{\textrm{QD1}}$,N$_{\textrm{QD2}}$) = (4,0)-(3,1) spin-blockaded charge anti-crossing. Supplementary Figs. \ref{fig:Appendix:Martin1}(c,d) show the magnetic field dependence of spin-orbit driven qubit rotations as a function of magnetic field applied along the [100] crystallographic direction. The spin-valley hot spot at 0.22 T indicates a valley splitting in this device of $\approx$ 25 $\mathrm{\mu eV}$. Electrical control of the hot spot driven qubit frequency is shown in Supplementary Figs. \ref{fig:Appendix:Martin1}(e,f), where a qubit drive frequency of 400 MHz is achieved.

\begin{figure*}[!]
	\centering
	\includegraphics[width=0.90\textwidth]{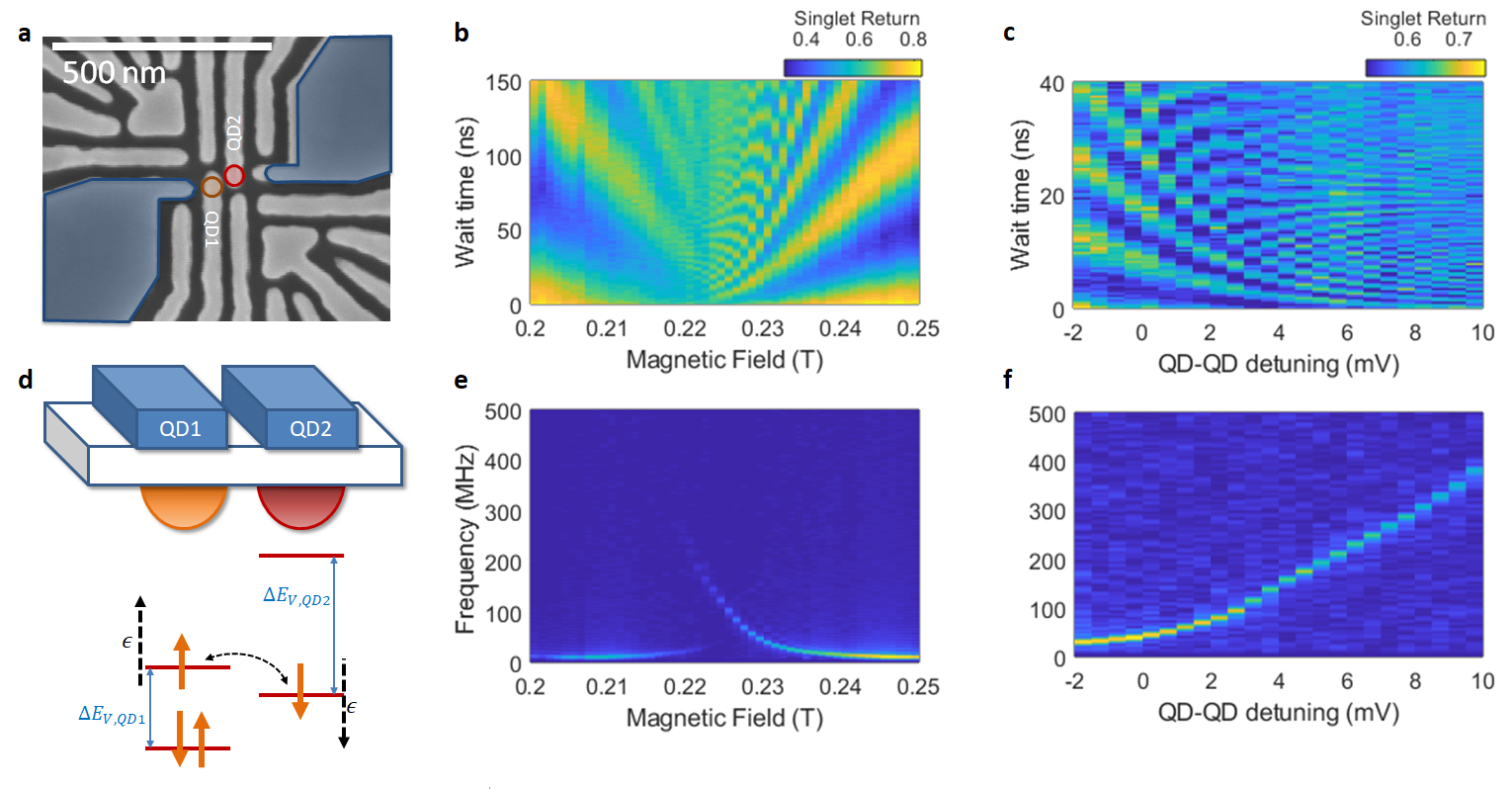}
	\caption{Spin-valley interaction in a $^{nat}$Si DQD device.(a) Scanning electron micrograph of the gate structure of a device similar to that measured. The overlaid regions indicate the estimated locations of electron accumulation (b) Change in singlet return as a function of $X$-rotation manipulation time as the magnetic field is varied along the [100] crystallographic direction. (c)  Change in singlet return as a function of $X$-rotation manipulation time as the QD-QD detuning is varied. (d) A cartoon representation of the electron spin filling in each QD in this device. (e) The FFT extracted rotation frequency as a function of magnetic field for the data in (b). (f) The FFT extracted rotation frequency as a function QD-QD detuning for the data in (c).}
	\label{fig:Appendix:Martin1}
\end{figure*}

\begin{figure*}[!]
	\centering
	\includegraphics[width=0.85\textwidth]{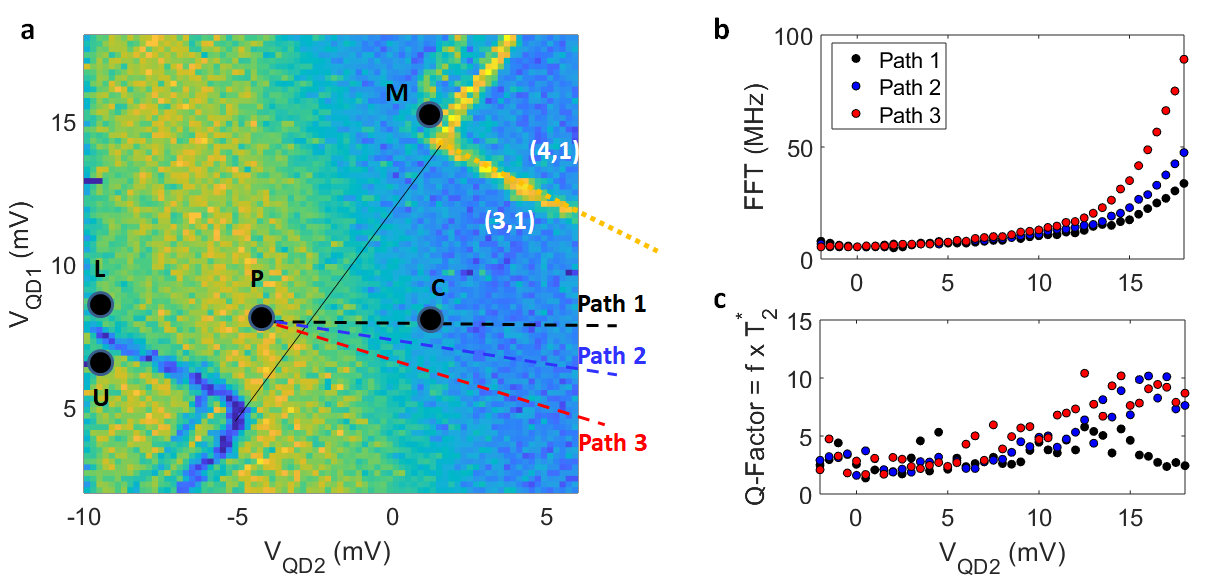}
	\caption{DQD tuning path. (a) A pulsed charge stability diagram for the (N$_{\textrm{QD1}}$,N$_{\textrm{QD2}}$) = (4,0)-(3,1) anti-crossing in Device B, showing the gradient of charge sensor current. The black circles represent the qubit reset (U), load (L), plunge (P), manipulation (C) and readout (M) points. The qubit manipulation point is varied in three experiments along three detuning paths (black, blue, and red dashed lines). (b) Extracted rotation frequency vs voltage applied to QD2 for the three paths. (c) The extracted frequency and dephasing times, $T_2^*$, give the Q-factors of the data for paths 1, 2, and 3.}
	\label{fig:Appendix:Martin2}
\end{figure*}

In Supplementary Fig. \ref{fig:Appendix:Martin2} we show the electrical control of the qubit frequency as a function of voltage applied to QD2 for QD-QD detunings along three separate paths. The different paths behave similarly, yet show differences in the plots. This suggests that while the vertical electric field influences the valley splitting of QD2, the voltage on both the QD1 and QD2 gates modify the the intervalley spin-orbit interaction.

\subsection{Valley splitting lever arm}
We find that the ability to electrically modulate the intervalley SOC is consistent with control of the valley splitting, $\Delta_{\mathrm{v}}$, through the applied gate voltages. Since the valley splitting plays an important role in dictating the magnetic field at which the polarized triplet state $T_{-}^{(1)}= \ket{\downarrow \downarrow^{(1)}}$ comes into resonance with the spin state $\ket{\downarrow \uparrow}$, we can use the voltage dependence of the qubit evolution frequency in the vicinity of the hot spot to probe the variation of valley splitting with gate voltage.

To do this, we first fit to the qubit frequency versus magnetic field data of Fig. \ref{fig:Exp:VoltageControl}(b) for Device A and \ref{fig:Appendix:Martin1}(e) for Device B, respectively.
We then fix these fit parameters and assume a linear variation of the valley splitting as a function of deviation of the gate voltage from the operating point at which the preceding measurements were taken,
\begin{equation}
    \Delta_{\mathrm{v}} = \Delta_{\mathrm{v}}^{0} + \lambda_{\mathrm{v}} (V-V_{0})
\end{equation}

\begin{figure*}[!]
	\centering
	\includegraphics[width=0.9\textwidth]{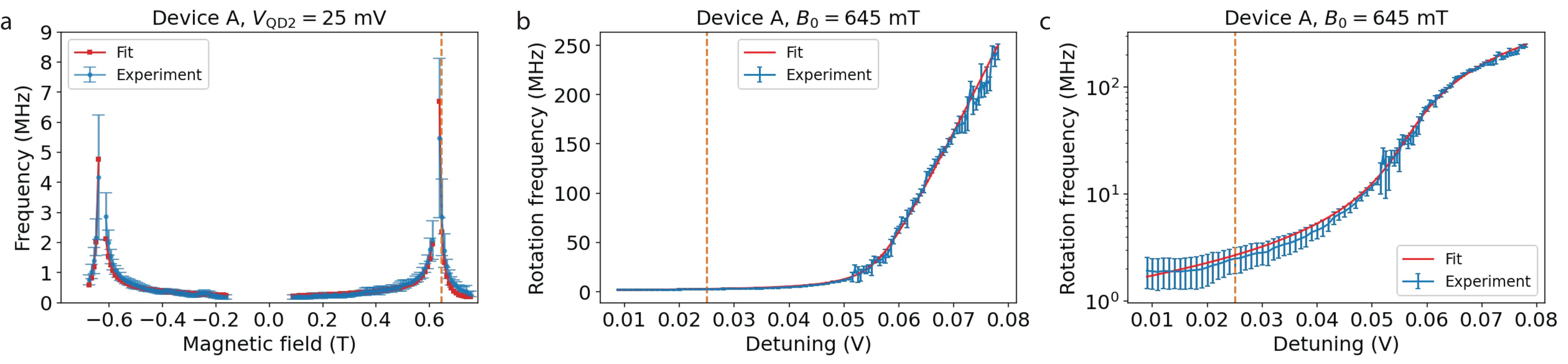}
	\caption{Model fits for Device A. Fits to magnetic field dependence (a) and gate voltage dependence (b and c, with same data plotted with frequency on linear and logarithmic scales for clarity, respectively). The vertical dashed line in each plot corresponds to the parameter held fixed in the other plot. Error bars represent $\pm \sigma$ for Gaussian fits to linewidths.}
	\label{fig:Appendix:DevA}
\end{figure*}

\begin{figure*}[!]
	\centering
	\includegraphics[width=0.6\textwidth]{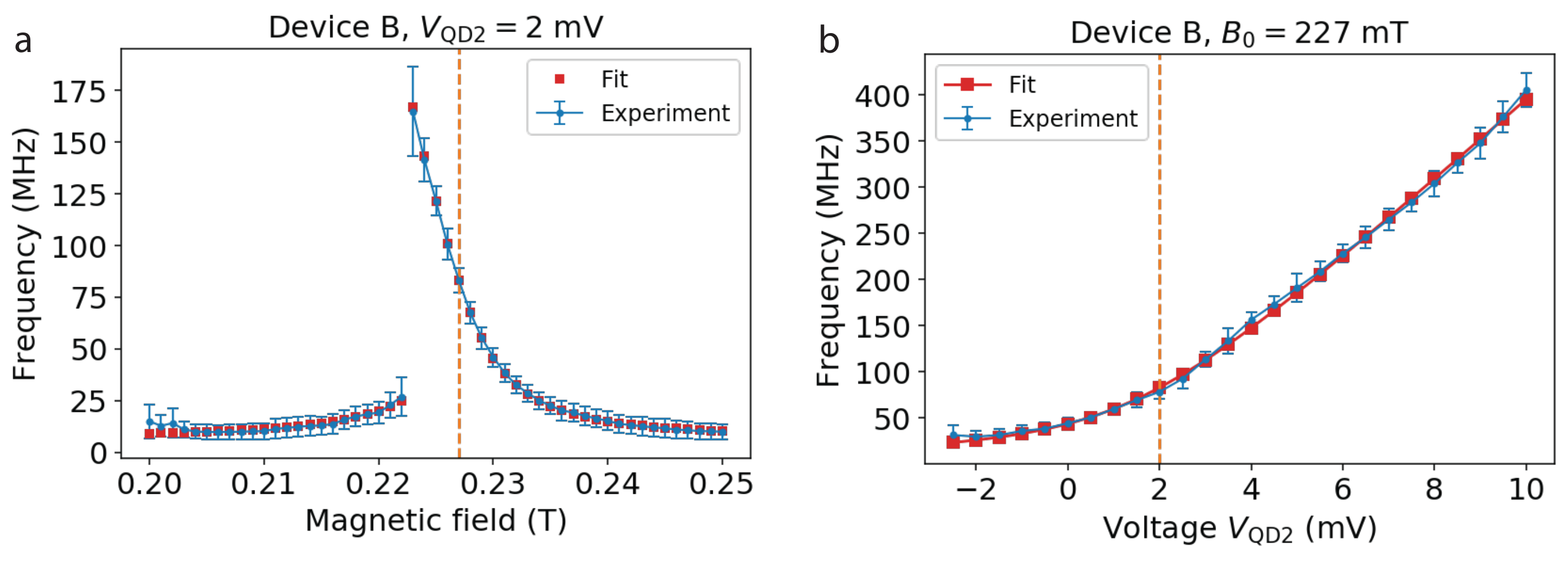}
	\caption{Model fits for Device B. Fits to magnetic field dependence (a) and gate voltage dependence (b). The vertical dashed line in each plot corresponds to the parameter held fixed in the other plot. Error bars represent $\pm \sigma$ for Gaussian fits to linewidths.}
	\label{fig:Appendix:DevB}
\end{figure*}

We show fits to the model parameters in Figs. \ref{fig:Appendix:DevA} and \ref{fig:Appendix:DevB}, with parameter estimates in Table \ref{tab:ModelParameters}. The reported uncertainties correspond to 95\% confidence intervals. Note that these measurements do not permit unambiguous determination of the valley-averaged $g$-factor $g_{*}$ of Eq. \ref{eq:HHotSpot}, so for the purpose of these parameter estimates we enforce $g_{*} = 2$. The results shown in Table \ref{tab:ModelParameters} are comparable with other results in silicon MOS QDs for valley splitting lever arms\cite{Yang2013,Gamble2016,Zhang2020} and intervalley SOC strengths\cite{Yang2013,Hao2014,HWang2017}.

\begin{table*}
    \centering
    \begin{tabular}{|c|c|c|}
    \hline
    & Device A & Device B  \\
    \hline \hline
    Intervalley SOC, $\gamma$ ($\mathrm{\mu eV}$) & $0.132 \pm 0.014$ & $0.28 \pm 0.05$ \\
    \hline
    Effective field gradient, $(10^{3}/h)\delta$ ($\mathrm{MHz/T}$) & $0.21 \pm 0.15$ & $0 \pm 15$ \\
    \hline
    Valley splitting, $\Delta_{\mathrm{v}}$ ($\mathrm{\mu eV}$) & $73.177 \pm 0.033$ & $26.39 \pm^{0.12}_{0.14}$\\
    \hline
    Valley splitting lever arm, $\lambda_{\mathrm{v}}$ ($\mathrm{\mu eV/V}$) & $46.25 \pm 0.85$ & $188 \pm 16$ \\
    \hline
    Residual exchange, $J_{31}$ ($\mathrm{neV}$)& $0 \pm 1.4$ & $0 \pm 60$ \\
    \hline
    \end{tabular}
    \caption{Model parameters for Device A (800 ppm $^{29}$Si device of the main text) and Device B ($^{\textrm{nat}}$Si device providing supporting independent measurements). The reported valley splitting for each device is that of the quantum dot associated with the measured spin-valley hot spot. The valley splitting lever arms correspond to collective variation of the dot gates (QD1,QD2) by ($-V$,$V$). Reported uncertainties are 95\% confidence intervals based on a $\chi^{2}$ analysis using the estimated linewidths. The parameter $J_{31}$ accounts for incomplete vanishing of the exchange coupling $J(\epsilon)$ in the large positive detuning regime in which these measurements were taken.}
    \label{tab:ModelParameters}
\end{table*}

\subsection{Impact of valley splitting}
\label{sec:Appendix:Challenges}

All silicon QD-based qubit architectures require sufficiently large valley splittings. Here, we briefly discuss a few of the implications of valley splitting on operating a singlet-triplet qubit or ensemble of singlet-triplet qubits using this hot-spot driven control scheme.

Since the applied magnetic field is assumed to be uniform across the sample, one requirement is that the hot spot must occur at similar magnetic field values for at least one dot of any given pair of quantum dots for all double dots in the ensemble. We have demonstrated here significant electrical control over the valley splitting, as indicated in Table \ref{tab:ModelParameters}. As long as the valley splittings for all relevant dots in the device are within the range of electrical tuning, in principle the hot spot driving scheme should be accessible. Given the measured valley splitting lever arm of 46 and 188 $\mathrm{\mu eV/V}$ for the isotopically purified and natural Si devices, respectively, we can estimate roughly how much valley splitting uniformity would be required. The range of voltage values over which we routinely drive the qubit is on the order of 100 $\mathrm{mV}$, which would correspond to a tunability of the valley splitting over a range of tens of $\mathrm{\mu eV}$. Hence, a uniformity of valley splitting within this range would presumably allow for hot spot operation for all qubits in the ensemble. Valley splitting has been shown to be tunable by a few hundred $\mu$eV in other MOS QD device geometries \cite{Yang2013,Gamble2016}, which would further reduce the constraint on uniformity.